\documentclass[aps,prd,reprint,nofootinbib,amsmath,amssymb,superscriptaddress]{revtex4-1} 
\usepackage{graphicx}   
\usepackage[
colorlinks=true,        
citecolor=blue,         
linkcolor=blue,         
urlcolor=blue           
]{hyperref}  
\usepackage{tabulary}
\usepackage{color}      
\usepackage{orcidlink}

\newcommand{\nc}{\newcommand*} 
\nc{\be}{\begin{equation}}
\nc{\ee}{\end{equation}}
\def\({\left(}
\def\){\right)}
\def\[{\left[}
\def\]{\right]}
\nc{\Eq}[1]{Eq.~\eqref{#1}}     
\nc{\Fig}[1]{Fig.~\ref{#1}}     
\nc{\Table}[1]{Table~\ref{#1}}  
\nc{\Sec}[1]{Sec.~\ref{#1}}     
\nc{\red}[1]{\textcolor{red}{#1}}

\begin{document}
\title{NANOGrav hints for first-order confinement-deconfinement phase transition\\
in different QCD-matter scenarios}

\author{Zu-Cheng Chen\orcidlink{0000-0001-7016-9934}}
\email{zuchengchen@hunnu.edu.cn}
\affiliation{Department of Physics and Synergetic Innovation Center for Quantum Effects and Applications, Hunan Normal University, Changsha, Hunan 410081, China}
\affiliation{Institute of Interdisciplinary Studies, Hunan Normal University, Changsha, Hunan 410081, China}
\affiliation{Department of Astronomy, Beijing Normal University, Beijing 100875, China}
\affiliation{Advanced Institute of Natural Sciences, Beijing Normal University, Zhuhai 519087, China}

\author{Shou-Long Li\orcidlink{0000-0002-7192-8717}}
\email{shoulongli@hunnu.edu.cn}
\affiliation{Department of Physics and Synergetic Innovation Center for Quantum Effects and Applications, Hunan Normal University, Changsha, Hunan 410081, China}
\affiliation{Institute of Interdisciplinary Studies, Hunan Normal University, Changsha, Hunan 410081, China}

\author{Puxun Wu\orcidlink{0000-0002-9188-7393}}
\email{pxwu@hunnu.edu.cn}
\affiliation{Department of Physics and Synergetic Innovation Center for Quantum Effects and Applications, Hunan Normal University, Changsha, Hunan 410081, China}
\affiliation{Institute of Interdisciplinary Studies, Hunan Normal University, Changsha, Hunan 410081, China}

\author{Hongwei Yu\orcidlink{0000-0002-3303-9724}}
\email{Corresponding author: hwyu@hunnu.edu.cn}
\affiliation{Department of Physics and Synergetic Innovation Center for Quantum Effects and Applications, Hunan Normal University, Changsha, Hunan 410081, China}
\affiliation{Institute of Interdisciplinary Studies, Hunan Normal University, Changsha, Hunan 410081, China}

\begin{abstract}
Recent observations from several pulsar timing array (PTA) collaborations have unveiled compelling evidence for a stochastic signal in the nanohertz band. This signal aligns remarkably with a gravitational wave (GW) background, potentially originating from the first-order color charge confinement phase transition. Distinct quantum chromodynamics (QCD) matters, such as quarks or gluons, and diverse phase transition processes thereof can yield disparate GW energy density spectra.
In this paper, employing the Bayesian analysis on the NANOGrav 15-year data set, we explore the compatibility with the observed PTA signal of the GW from phase transitions of various QCD matter scenarios in the framework of the holographic QCD.
We find that the PTA signal can be effectively explained by the GW  from the confinement-deconfinement phase transition of pure quark systems in a hard wall model of the holographic QCD where the bubble dynamics, one important source of the GWs, is of the Jouguet detonations.
Notably, our analysis decisively rules out the plausibility of the pure gluon QCD-matter scenario and the non-runaway bubble dynamics model for the phase transition in explaining the observed PTA signal.

\end{abstract}

\maketitle

\section{Introduction}
The detections of gravitational waves (GWs) from compact binary coalescences by ground-based detectors~\cite{LIGOScientific:2018mvr,LIGOScientific:2020ibl,LIGOScientific:2021djp} in the culmination of a decades-long quest confirm directly a prediction of Einstein based on his classical theory of general relativity.
These detections not only have been utilized to test general relativity in the strong-field regime~\cite{LIGOScientific:2019fpa,LIGOScientific:2020tif,LIGOScientific:2021sio}, but also have allowed us to explore the properties of the broader population of GW sources~\cite{LIGOScientific:2018jsj,LIGOScientific:2020kqk,Chen:2021nxo,KAGRA:2021duu,Chen:2022fda}. 
Despite the great achievement of LIGO-Virgo-KAGRA in detecting individual merger events, the quest to detect a stochastic GW background (SGWB), another type of GW source, remains an ongoing pursuit. 
By inspecting the spatially correlated fluctuations induced by an SGWB on the arrival times of radio pulses from an array of pulsars~\cite{1978SvA....22...36S,Detweiler:1979wn,1990ApJ...361..300F}, we can use a pulsar timing array (PTA) as a valuable tool to search for the SGWB in the nanohertz frequency band.
Recently, several PTA collaborations, including the NANOGrav~\cite{NANOGrav:2023gor,NANOGrav:2023hde}, the PPTA~\cite{Zic:2023gta,Reardon:2023gzh}, the EPTA (in combination with the InPTA)~\cite{EPTA:2023sfo,Antoniadis:2023ott}, and the CPTA~\cite{Xu:2023wog}, have independently detected a stochastic signal shared by all pulsars. Furthermore, these PTAs found evidence, with varying levels of significance, which supports the Hellings-Downs~\cite{Hellings:1983fr} spatial correlations and suggests the SGWB origin of the signal.
While various potential sources exist within the PTA window~\cite{Chen:2019xse,Vagnozzi:2020gtf,Chen:2021wdo,Sakharov:2021dim,Benetti:2021uea,Chen:2022azo,Ashoorioon:2022raz,PPTA:2022eul,Chen:2023uiz,IPTA:2023ero,Dandoy:2023jot,Madge:2023cak,Bi:2023ewq,Wu:2023rib,Wu:2023dnp}, the nature of this signal as to whether it arises from astrophysical or cosmological phenomena remains a subject of extensive investigation~\cite{NANOGrav:2023hvm,Antoniadis:2023xlr,Athron:2023mer,Niu:2023bsr,Bi:2023tib,Wang:2023sij,Liu:2023pau,Vagnozzi:2023lwo,Fu:2023aab,Han:2023olf,Li:2023yaj,Franciolini:2023wjm,Shen:2023pan,Kitajima:2023cek,Franciolini:2023pbf,Addazi:2023jvg,Cai:2023dls,Inomata:2023zup,Murai:2023gkv,Li:2023bxy,Anchordoqui:2023tln,Liu:2023ymk,Figueroa:2023zhu,Yi:2023mbm,Zhao:2023joc,Wu:2023hsa,Bian:2023dnv,Geller:2023shn,Antusch:2023zjk,Ye:2023xyr,HosseiniMansoori:2023mqh,Jin:2023wri,Zhang:2023nrs,Choudhury:2023kam,Yi:2023npi,Gorji:2023sil,Das:2023nmm,Ellis:2023oxs,Balaji:2023ehk,Maji:2023fhv,Bhaumik:2023wmw,Basilakos:2023xof,Huang:2023chx,Jiang:2023gfe,Wang:2023len,DiBari:2023upq,Aghaie:2023lan,Harigaya:2023pmw,Kawai:2023nqs,Salvio:2023ynn,Lozanov:2023rcd,Choudhury:2023fwk,Cang:2023ysz,Liu:2023hpw,Lazarides:2023ksx,Choudhury:2023fjs,Fei:2023iel,Chen:2024fir,You:2023rmn,Yi:2023tdk}.

One potential source for this signal is the SGWB generated by a first-order phase transition in the early Universe, occurring as the temperature decreases to the scale of quantum chromodynamics (QCD)~\cite{He:2023ado,Zu:2023olm,Abe:2023yrw,Ghosh:2023aum,Gouttenoire:2023bqy,Bigazzi:2020avc,Li:2021qer}.
Here, the first-order cosmological phase transition refers to the transitions of the Universe from a metastable vacuum state to a stable vacuum state through quantum tunneling or thermal fluctuations via the nucleation of bubbles in a sea of metastable phase. These bubbles expand and eventually collide with each other.  The phase transition processes, especially the collision of the bubble walls and the subsequent plasma phenomena like sound waves and magnetohydrodynamic (MHD) turbulence, which are closely related to the dynamics of the bubble models, can lead to a significant SGWB~\cite{Kosowsky:1991ua,Kosowsky:1992rz,Kosowsky:1992vn,Kamionkowski:1993fg,Caprini:2007xq,Hindmarsh:2013xza,Giblin:2013kea,Giblin:2014qia,Kosowsky:2001xp,Caprini:2006jb,Kahniashvili:2008pf,Kahniashvili:2008pe,Kahniashvili:2009mf,Kisslinger:2015hua,Huber:2008hg,Caprini:2009yp,Caprini:2015zlo,Schwaller:2015tja,Hindmarsh:2015qta,Ellis:2018mja,Ellis:2020awk,Morgante:2022zvc,Lewicki:2022pdb}.
The model for the bubble collision, which is one important source of the GWs generated in the phase transition processes, may encompass both the Jouguet detonation scenario~\cite{Steinhardt:1981ct,Caprini:2015zlo,Kamionkowski:1993fg,Nicolis:2003tg,Espinosa:2010hh,Hindmarsh:2015qta}, characterized by strong detonations where the phase transition proceeds at the speed of sound, and the non-runaway bubble scenario~\cite{Espinosa:2010hh,Caprini:2015zlo,Hindmarsh:2015qta,Kamionkowski:1993fg}, involving a more moderate transition pace without extreme velocities.

 Specifically, the Universe may undergo, as the temperature decreases to the QCD scale, a first-order confinement-deconfinement phase transition, transitioning from a quark-gluon plasma (QGP) phase to a hadron phase. The transition temperatures can slightly vary for different QCD-matter plasma~\cite{He:2023ado,Zu:2023olm,Abe:2023yrw,Ghosh:2023aum,Gouttenoire:2023bqy,Li:2021qer}, such as pure quark or pure gluon systems. 
The diverse dynamics of the bubble evolution and the temperatures at which phase transitions occur can both imprint variations in the power spectrum of the SGWB generated by the phase transitions. Despite these variations, the frequencies of the SGWB consistently fall within the observable frequency range of PTAs. 
Therefore, we can use the PTA data to study the specific scenarios of the cosmic evolution at the QCD scale.

In a previous study~\cite{Li:2021qer} utilizing NANOGrav 12.5-year data set, it was suggested that the PTA signal could be potentially generated by the first-order confinement-deconfinement transitions of heavy static quarks plasma with a zero baryon chemical potential or quarks with a finite baryon chemical potential in both the Jouguet detonation and non-runaway bubble dynamics models. 
In this paper, by using Bayesian analysis on the NANOGrav 15-year data set, we delve deeper into exploring the compatibility with the PTA signal of the GWs resulting from phase transitions in various scenarios of QCD-matter plasma.
We examine three scenarios: (i) heavy static quarks at zero baryon chemical potential, (ii) quarks with a finite baryon chemical potential, and (iii) a pure gluon system.
We show that the PTA signal can be effectively explained by the GW from the phase transition of pure quark systems within the hard wall model of the holographic QCD in which the bubble dynamics is of Jouguet detonations, irrespective of the chemical potential's value. Notably, the PTA signal is incompatible with the GW generated by the phase transition of a pure gluon plasma and the non-runaway bubble dynamics model.

\begin{table*}[tbp]
  \begin{center}
  \begin{tabular}{c|cccc}
  \hline\hline
  Model & QCD matter & Holographic QCD-like model & $\alpha$ & $T_*$ \\ 
   \hline
  $S_1$ & Heavy static quarks with a zero chemical potential & Hard wall & $13.5$ & $122$\,MeV ~\cite{Herzog:2006ra,Ahmadvand:2017xrw}\\    
   $S_2$ & Heavy static quarks with a zero chemical potential & Soft wall & $4.27$ & $191$\,MeV~\cite{Herzog:2006ra,Ahmadvand:2017xrw}\\ 
  $S_3$ & Quarks with a finite chemical potential & Hard wall & $32.2$ & $112$\,MeV~\cite{Ahmadvand:2017tue}\\
   $S_4$ & Quarks with a finite chemical potential & Soft wall & $4.56$ & $192$\,MeV~\cite{Ahmadvand:2017tue}\\
  $S_5$ & Pure gluons & \hspace{2mm}Quenched dynamical holographic QCD\hspace{2mm} & \hspace{2mm}$0.611$\hspace{2mm} & $255$\,MeV~\cite{Chen:2017cyc} \\ 
   \hline
  \end{tabular}    
  \caption{\label{tab1}The ratio of the vacuum energy density $\alpha$ and critical temperature $T_*$ from five holographical QCD-like models.}
  \end{center}
\end{table*}

\section{GWs from first-order QCD phase transitions} 
The spectrum of an isotropic SGWB is characterized by the dimensionless GW energy density parameter $\rho_{\mathrm{GW}}$  per logarithm frequency $f$~\cite{Allen:1997ad}
\be
\Omega_{\mathrm{GW}}(f)=\frac{1}{\rho_c} \frac{d \rho_{\mathrm{GW}}}{d \ln f},
\ee 
where $\rho_c=3 H_0^2 / 8 \pi G$ is the critical energy density of the Universe in which $H_0$ is the Hubble constant. 
As mentioned earlier, the total energy density of GWs, $\rho_{\mathrm{GW}}$, mainly arises from three processes: collisions of vacuum bubble walls, and the subsequent sound waves and the MHD turbulence in the plasma~\cite{Kosowsky:1991ua,Kosowsky:1992rz,Kosowsky:1992vn,Kamionkowski:1993fg,Caprini:2007xq,Hindmarsh:2013xza,Giblin:2013kea,Giblin:2014qia,Kosowsky:2001xp,Caprini:2006jb,Kahniashvili:2008pf,Kahniashvili:2008pe,Kahniashvili:2009mf,Kisslinger:2015hua,Huber:2008hg,Caprini:2009yp,Caprini:2015zlo,Hindmarsh:2015qta}.
So, we have
\be
h^2 \Omega_{\mathrm{GW}}(f) \simeq h^2 \Omega_{\textup{en}} (f) + h^2 \Omega_{\textup{sw}} (f) + h^2 \Omega_{\textup{tu}} (f) \,, \label{GW}
\ee
where $h \equiv H_0/(100\,\mathrm{km}\,\mathrm{s}^{-1}\,\mathrm{Mpc}^{-1}) = 0.674$~\cite{Planck:2018vyg} is the dimensionless Hubble constant, and~\cite{Huber:2008hg,Caprini:2009yp,Caprini:2015zlo,Hindmarsh:2015qta}
\begin{align}
h^2 \Omega_{\textup{en}}  &\!=\! 3.6 \times 10^{-5}\! \left(\!\frac{H_\ast}{\beta}\!\right)^{\!\!2}\!\!\left(\!\frac{\kappa_1 \alpha}{1+\alpha}\!\right)^{\!\!2}\!\! \left(\!\frac{10}{g_\ast}\!\right)^{\!\!\frac13}\!\!\left(\!\frac{0.11 v_w^3}{0.42+v_w^2}\!\right)\!\! S_{\textup{en}}, \label{eq:Omega1}\\
h^2 \Omega_{\textup{sw}} &\!=\! 5.7 \times 10^{-6}\! \left(\frac{H_\ast}{\beta} \right)\!\! \left(\frac{\kappa_2 \alpha}{1+\alpha}\right)^{\!2}\! \left(\frac{10}{g_\ast} \right)^{\!\!\frac13}  v_w S_{\textup{sw}}, \label{eq:Omega2}\\
h^2 \Omega_{\textup{tu}}  &\!=\! 7.2 \times 10^{-4}\! \left(\frac{H_\ast}{\beta} \right)\!\left(\frac{\kappa_3 \alpha}{1+\alpha} \right)^{\!2}\!\! \left(\frac{10}{g_\ast} \right)^{\!\frac13}\! v_w S_{\textup{tu}}, \label{eq:Omega3}
\end{align}
are the contributions respectively from collisions of the bubble walls and the sound waves, as well as from the turbulence. Here, $\alpha$ is the ratio of the vacuum energy density to the radiation energy density, $\beta$ is the inverse time duration of the phase transition, and $v_w$ denotes the velocity of the bubble wall. Meanwhile, $g_\ast $ and $ H_\ast$ correspond to the number of active degrees of freedom and the Hubble parameter at the time when GWs are produced.
We set $g_\ast =10$ and $\beta/H_\ast =10$~\cite{Ahmadvand:2017xrw,Ahmadvand:2017tue,Chen:2017cyc} at the QCD scale. The spectral shapes of GWs are expressed by numerical fits as~\cite{Huber:2008hg,Caprini:2015zlo,Hindmarsh:2015qta,Caprini:2009yp} 
\begin{align}
S_{\textup{en}} &= \frac{3.8 \left(f/f_\mathrm{en} \right)^{2.8}}{1+2.8 \left(f/f_\mathrm{en} \right)^{3.8}}, \\
S_{\textup{sw}} &= \left(f/f_\mathrm{sw}\right)^3 \left[\frac{7}{4 +3 \left(f/f_\mathrm{sw}\right)^2} \right]^\frac{7}{2},\\
S_{\textup{tu}}  &= \frac{\left(f/f_\mathrm{tu}\right)^3 }{\left[1+ \left(f/f_\mathrm{tu}\right) \right]^{\frac{11}{3}} \left(1 +  8 \pi f/h_{\ast}\right)},    
\end{align}
where $h_{\ast} = 11.2\, \mathrm{nHz}\, [T_\ast/(100\,\mathrm{MeV})] (g_\ast/10)^{1/6}$ is the Hubble rate at $T_\ast$, the temperature of the thermal bath when GWs are generated. We assume that the GWs are generated promptly after the phase transition, hence $T_\ast$ approximates the critical phase transition temperature. The peak frequencies are given by~\cite{Kamionkowski:1993fg,Caprini:2015zlo,Caprini:2009yp}
\begin{align}
f_{\textup{en}} &\!=\! 11.2\, \textup{nHz}\! \left(\!\frac{0.62}{1.8 -0.1 v_w +v_w^2}\!\right)\!\!\! \left(\!\frac{\beta}{H_\ast}\!\right) \! \!\!\left(\!\frac{T_\ast}{100 \textup{MeV}}\!\right) \!\! \left(\!\frac{g_\ast}{10}\!\right)^{\!\frac16}, \\
f_{\textup{sw}} &\!=\! 12.9\, \textup{nHz}\! \left(\frac{1}{v_w}\right) \left(\frac{\beta}{H_\ast}\right) \left(\frac{T_\ast}{100 \textup{MeV}}\right) \left(\frac{g_\ast}{10} \right)^{\frac16}, \\
f_{\textup{tu}} &\!=\! 18.4\, \textup{nHz}\! \left(\frac{1}{v_w}\right) \left(\frac{\beta}{H_\ast}\right) \left(\frac{T_\ast}{100 \textup{MeV}}\right) \left(\frac{g_\ast}{10} \right)^{\frac16}.
\end{align}
The coefficients $\kappa_1 , \kappa_2 $, and $ \kappa_3 $ in Eqs.~(\ref{eq:Omega1}-\ref{eq:Omega3}) are model dependent. These coefficients represent fractions of the vacuum energy converted to the kinetic energy of the bubbles, bulk fluid motion, and the MHD turbulence, respectively. In this work, we examine two bubble scenarios: Jouguet detonations and non-runaway bubbles. For Jouguet detonations~\cite{Steinhardt:1981ct,Caprini:2015zlo,Kamionkowski:1993fg,Nicolis:2003tg,Espinosa:2010hh,Hindmarsh:2015qta}, we have $\kappa_1 = (0.715 \alpha +0.181 \sqrt{\alpha})/(1+ 0.715 \alpha)$, $ \kappa_2 = \sqrt{\alpha}/(0.135 + \sqrt{\alpha +0.98})$, $\kappa_3 = 0.05 \kappa_2$, and $v_w = (\sqrt{1/3} +\sqrt{\alpha^2 +2 \alpha/3})/( 1+ \alpha)$.
For non-runaway bubbles~\cite{Espinosa:2010hh,Caprini:2015zlo,Hindmarsh:2015qta,Kamionkowski:1993fg}, we have $\kappa_1 = 0$, $\kappa_2 = \alpha/(0.73 +0.083 \sqrt{\alpha} +\alpha)$, $\kappa_3 = 0.05 \kappa_2$, and $v_w = 0.95$.

\begin{figure*}[t]
    \centering
    \includegraphics[width=0.4\linewidth]{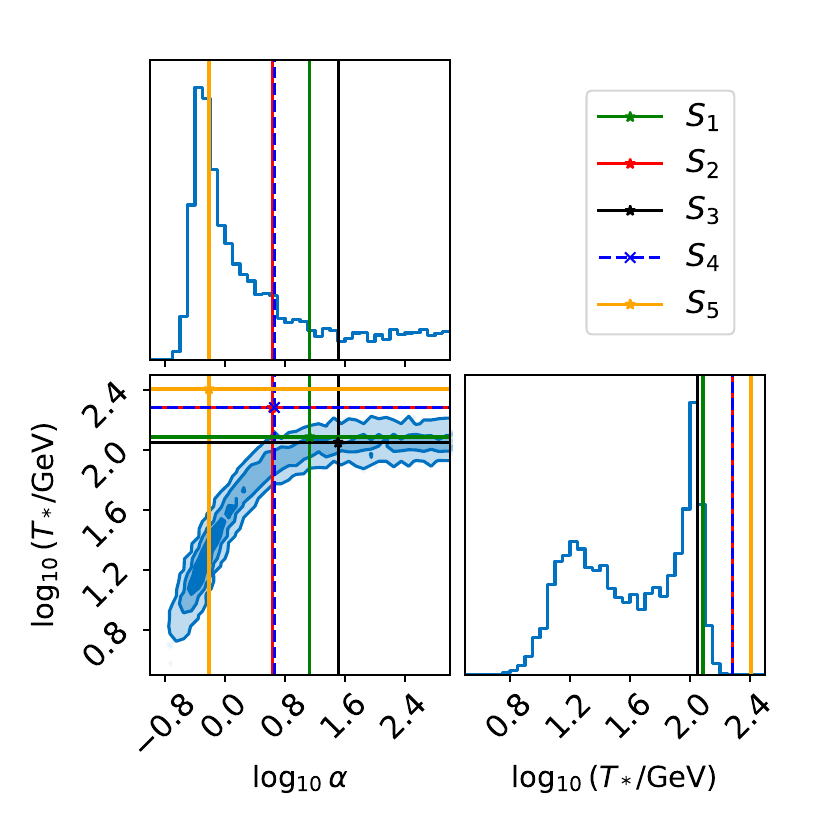}
    \includegraphics[width=0.56\linewidth]{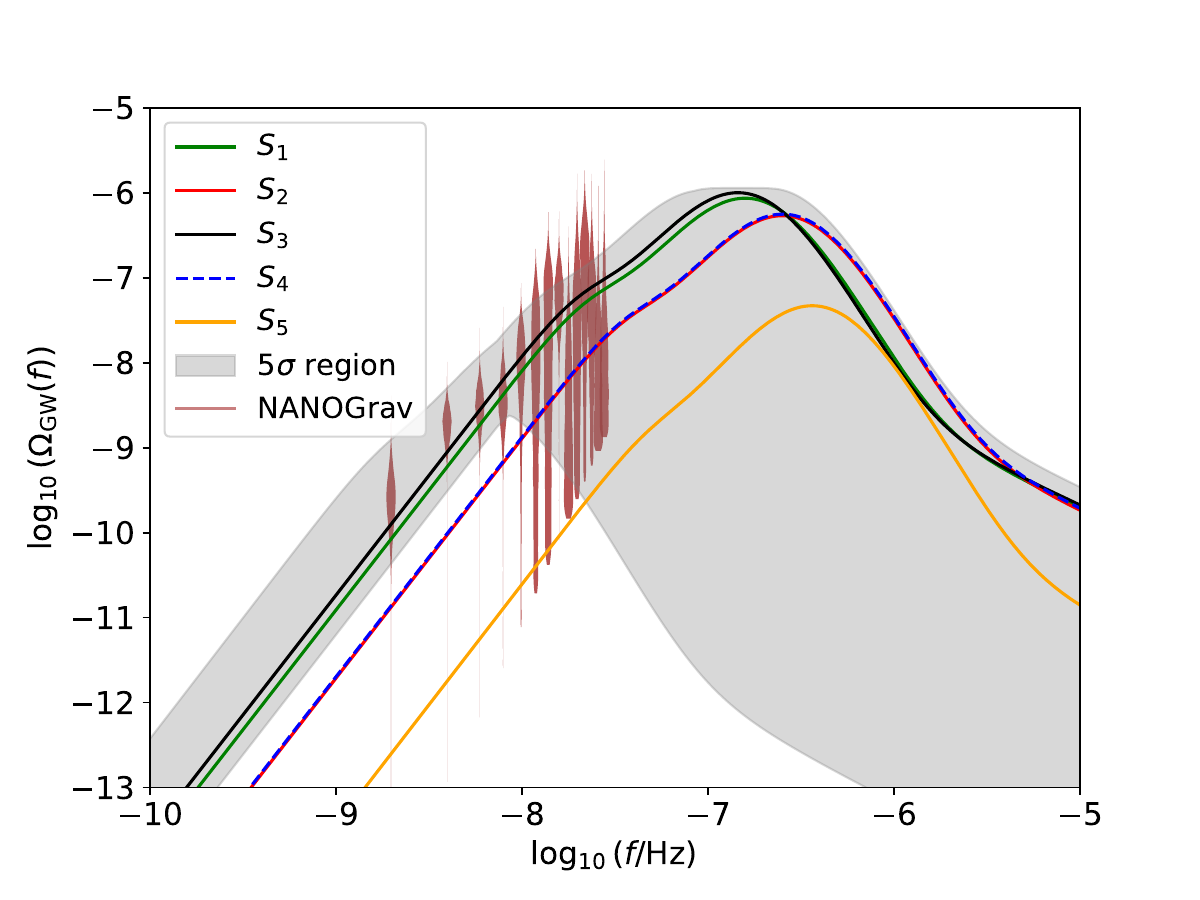}
    \caption{\label{posts_Jouguet}\textbf{Left panel:} Bayesian posteriors for model parameters $\alpha$ and $T_*$ in the Jouguet detonation bubble scenario, using the NANOGrav 15-year data set. We show the $1\sigma$, $2\sigma$, and $5\sigma$ contours in the two-dimensional plot. The five holographical models are also indicated in the parameter space. \textbf{Right panel:} Posterior predictive distributions for the GW spectrum derived from the NANOGrav 15-year data set. The grey region represents the $5\sigma$ confidence interval, while the brown violins depict free spectrum data from the NANOGrav 15-year data set. Additionally, We show the GW spectra from the QCD-matter confinement-deconfinement phase transition in the Jouguet detonation bubble case from five holographical models.
    }
\end{figure*}  

The expression in \Eq{GW} relies entirely on $\alpha$ and $T_\ast$, which are related to the QCD matter scenario and the phase transition under consideration. 
Normally, one may expect to determine the specific values of $\alpha$ and $T_*$ by the standard lattice QCD simulations.
However, these simulations encounter limitations due to the sign problem, yielding very few applicable results~\cite{Takeda:2011vd,Fromm:2011qi,Ding:2017giu}.
To circumvent these difficulties, one then adopts the Anti-de Sitter (AdS)/QCD framework that is established upon the AdS/conformal field theory (CFT) correspondence principle~\cite{Maldacena:1997re,Gubser:1998bc,Witten:1998qj}.
This framework offers novel insights into non-perturbative hadron dynamics through the dual gravitational field~\cite{Erlich:2005qh}. Specifically, it interprets the first-order confinement-deconfinement phase transitions via Hawking-Page (HP) transitions~\cite{Hawking:1982dh} in five-dimensional spacetime within AdS/QCD models~\cite{Herzog:2006ra}. In this framework, the high-temperature quark-gluon plasma corresponds to the AdS black hole, while the low-temperature hadron phase aligns with thermal AdS space.
The generation of GWs from the first-order QCD phase transition has been studied within this framework~\cite{Randall:2006py}, considering systems with heavy static quarks under both zero~\cite{Ahmadvand:2017xrw} and finite baryon chemical potentials~\cite{Ahmadvand:2017tue,Ahmadvand:2020fqv}, as well as in pure gluon systems~\cite{Chen:2017cyc}, and employing various AdS/QCD models. 
Following~\cite{Li:2021qer}, we explore the three types of QCD matter systems across five distinct holographic models: (i) heavy static quarks with a zero baryon chemical potential in hard-wall model $S_1$ and (ii) soft-wall model $S_2$, (iii) quarks with a finite baryon chemical potential in hard-wall model $S_3$ and (iv)
soft-wall model $S_4$, and (v) pure gluons in the quenched dynamical holographic model $S_5$. 
For a specific model, one can obtain the value of $\alpha$ and $T_\ast$ via some holographic techniques. 
\Table{tab1} summarizes the corresponding values, and we refer  our readers to~\cite{Herzog:2006ra,Ahmadvand:2017tue,Ahmadvand:2017xrw,Chen:2017cyc} for detailed calculations.

\begin{figure*}[t]
    \centering
    \includegraphics[width=0.4\linewidth]{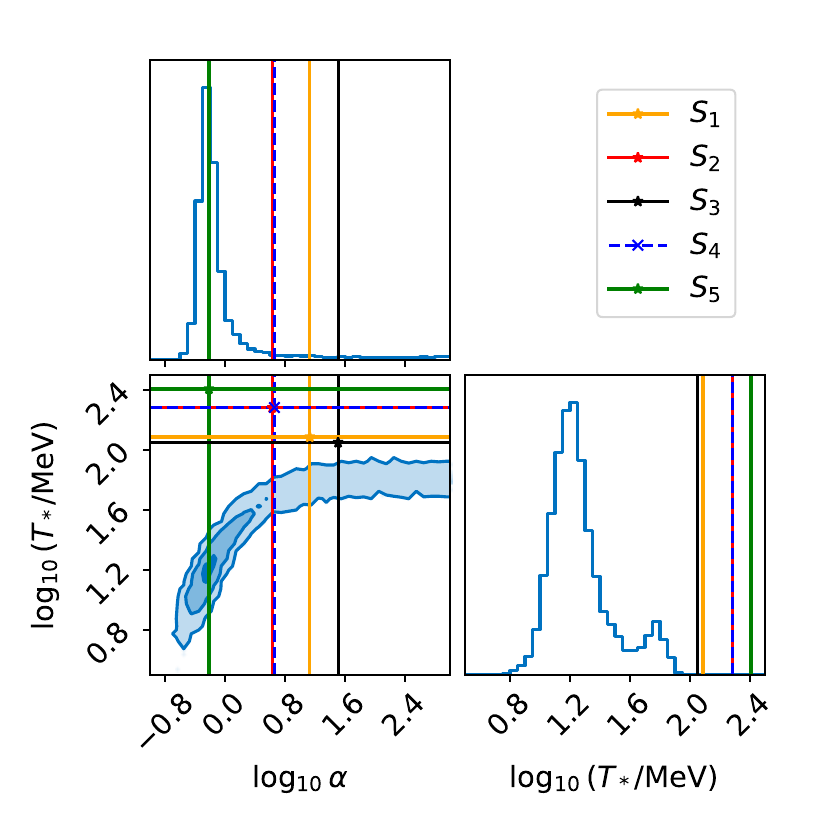}
    \includegraphics[width=0.56\linewidth]{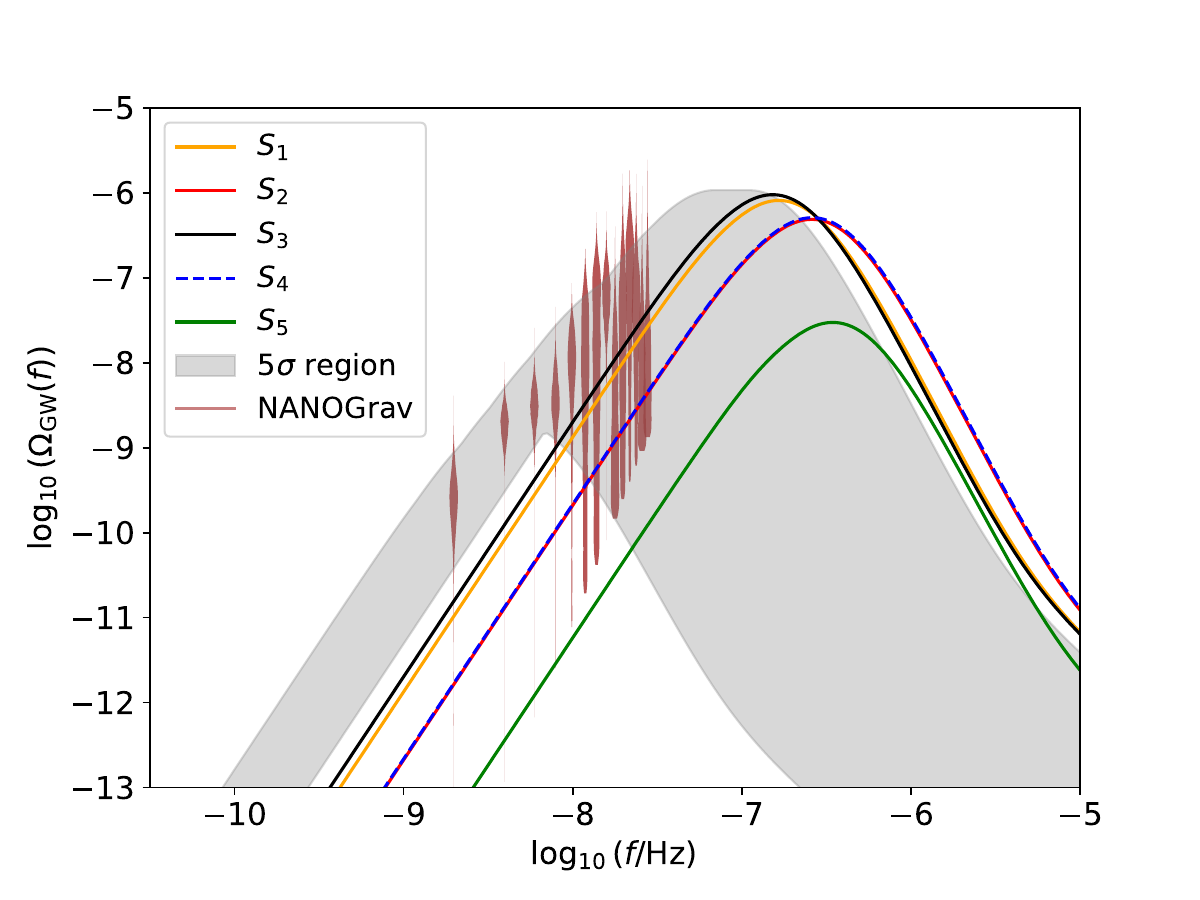}
    \caption{\label{posts_nonrunaway} Same as \Fig{posts_Jouguet} but for the non-runaway bubble case.}
\end{figure*}  

\section{Data analyses and results}
In this work, we analyze the NANOGrav 15-year data set~\cite{NANOGrav:2023hde} to estimate model parameters $\alpha$ and $T_*$. Specifically, we utilize the free spectrum amplitudes derived by NANOGrav, considering the full spatial correlations following the Hellings-Downs pattern. The sensitivity of a PTA's observation begins at a frequency of $1/T_{\mathrm{obs}}$, with $T_{\mathrm{obs}} = 16.03\,\mathrm{yr}$ representing the observational time span of NANOGrav.
Following NANOGrav~\cite{NANOGrav:2023gor}, we employ $14$ frequencies for Bayesian inference.

We start with the posterior data for time delay, $d(f)$, released by NANOGrav~\cite{NANOGrav:2023gor}. The power spectrum $S(f)$ is linked to the time delay $d(f)$ through
\begin{equation}
S(f) = d(f)^2\, T_{\mathrm{obs}}.
\end{equation}
By using the time delay data, we can further compute the GW energy density of the free spectrum as
\begin{equation}\label{hatomega}
\hat{\Omega}_{\mathrm{GW}}(f)=\frac{2 \pi^2}{3 H_0^2} f^2 h_c^2(f) = \frac{8\pi^4}{H_0^2} T_{\mathrm{obs}} f^5 d^2(f),
\end{equation}
where $h_c(f)$ is the characteristic strain defined by
\begin{equation}
h_c^2(f)=12 \pi^2 f^3 S(f).
\end{equation}
For each observed frequency $f_i$, we employ the posteriors of $\hat{\Omega}_{\mathrm{GW}}(f_i)$ obtained in \Eq{hatomega} to estimate the corresponding kernel density, $\mathcal{L}_i(\hat{\Omega}_{\mathrm{GW}})$. Consequently, the total log-likelihood is the summation of individual log-likelihoods given by~\cite{Moore:2021ibq,Lamb:2023jls,Liu:2023ymk,Wu:2023hsa,Jin:2023wri,Liu:2023pau}
\begin{equation}
\ln \mathcal{L}(\Lambda) = \sum_{i=1}^{66} \ln \mathcal{L}_i(\Omega_{\mathrm{GW}}(f_i, \Lambda)),
\end{equation}
where $\Lambda\equiv \{\alpha, T_*\}$ represents the two model parameters.
We utilize the \texttt{dynesty}~\cite{Speagle:2019ivv} sampler available within the \texttt{Bilby}~\cite{Ashton:2018jfp,Romero-Shaw:2020owr} package to explore the parameter space. To evaluate the model fits, we calculate the Bayesian Information Criterion (BIC)~\cite{Schwarz_1978}, defined as
\begin{equation}
\mathrm{BIC} = k \ln n - 2 \ln \hat{\mathcal{L}},
\end{equation}
where $n=66$ corresponds to the number of data points in the frequency space, $k$ is the number of parameters in the model, and $\hat{\mathcal{L}}$ is the maximum likelihood. The BIC facilitates the comparison of two competing models~\cite{kass1995bayes}: if the difference in BICs between the two models is $0-2$, it constitutes `weak' evidence in favor of the model with the smaller BIC; a difference in BICs between 2 and 6 indicates `positive' evidence; a difference in BICs between 6 and 10 signifies `strong' evidence; and a difference in BICs greater than 10 indicates `very strong' evidence in favor of the model with a smaller BIC.

For the model parameters, we sample them in logarithm space. Specifically, we employ priors such that $\log_{10} \alpha$ is uniformly distributed in $[-1, 3]$, and $\log_{10} (T_*/\mathrm{MeV})$ is uniformly distributed in $[0.5, 2.5]$.
For the Jouguet detonation case, the NANOGrav data yields $\alpha=1.22^{+351}_{-0.86}$ and $T_* = 42^{+76}_{-31}\,\mathrm{MeV}$. Unless specified otherwise, we report the median value and $90\%$ equal-tail uncertainties for each parameter.
\Fig{posts_Jouguet} shows the posteriors of model parameters and the corresponding GW energy density in the Jouguet detonation case. 
Notably, the NANOGrav 15-year data set is in agreement with the $S_1$ and $S_3$ models, suggesting that the confinement-deconfinement phase transition in pure quark systems (specifically, cases involving heavy static quarks with a zero baryon chemical potential and quarks with a finite baryon chemical potential) can serve as the cosmological sources of the PTA signal in the Jouguet detonation scenario. Furthermore, both $S_1$ and $S_3$ belong to the hard wall holographic model, thereby excluding the soft wall holographic model. In contrast, the confinement-deconfinement transition in the pure gluon system is unlikely the source of the NANOGrav signal.
Since the critical phase transition temperatures for the cases of a finite chemical potential and a zero chemical potential are very close, the chemical potential exerts negligible influence on the power spectrum of GWs. Thus, the current NANOGrav observation indicates that quark confinement dominates the cosmological QCD transition. 
For the non-runaway bubble case, the posteriors are $\alpha=0.63^{+35}_{-0.23}$ and $T_* = 17^{+42}_{-6.6}\,\mathrm{MeV}$.
\Fig{posts_nonrunaway} displays the results for the non-runaway bubble case. 
It is evident that all five holographic models are ruled out by the NANOGrav 15-year data set at a $5\sigma$ confidence level. Therefore, the NANOGrav data decisively excludes the non-runaway bubble case.

In \Table{tab:BIC1}, we present the differences in BIC among the ten QCD-like models with $\beta/H_* = 10$. Remarkably, the $S_1$ and $S_3$ models within the Jouguet bubble scenario exhibit superior performance compared to all other models. Furthermore, there is `positive' evidence supporting the preference for the $S_3$ model over the $S_1$ model in this scenario. Extending the parameter $\beta/H_*$ to a larger value, such as $\beta/H_* = 15$, as shown in \Table{tab:BIC2}, the $S_1$ and $S_3$ models in the Jouguet bubble scenario maintain their superior performance over other models, with `very strong' evidence favoring the $S_3$ model over the $S_1$ model. Therefore, a larger value of $\beta/H_*$ does not significantly alter our findings.

\begin{table}[tbp]
  \begin{center}
  \begin{tabular}{c|ccccc}
  \hline\hline
   & $S_1$ & $S_2$ & $S_3$ & $S_4$ & $S_5$ \\ 
   \hline
   Jouguet &  0 & 42.5 & -3.1 & 41.2 & 63.4 \\
   non-runaway & 43.0 & 60.2 & 34.0 & 60.1 & 59.2\\
   \hline 
  \end{tabular}    
  \caption{\label{tab:BIC1}The difference of BIC among the ten QCD-like models with fixing $\beta/H_* = 10$. We have set the $S_1$ from the Jouguet bubble case as the fiducial model.}
  \end{center}
\end{table}

\begin{table}[tbp]
  \begin{center}
  \begin{tabular}{c|ccccc}
  \hline\hline
   & $S_1$ & $S_2$ & $S_3$ & $S_4$ & $S_5$ \\ 
   \hline
   Jouguet &  0 & 17.5 & -12.9 & 17.4 & 17.0 \\
   non-runaway & 16.6 &  18.6 & 19.9 & 18.6 & 16.3 \\
   \hline 
  \end{tabular}    
  \caption{\label{tab:BIC2}Same as \Table{tab:BIC1}, but fixing $\beta/H_* = 15$.}
  \end{center}
\end{table}

\section{Summary and discussion}
In this paper, we firstly demonstrate that the stochastic signal detected in the NANOGrav 15-year data set can be successfully attributed to an SGWB arising from the first-order cosmological confinement-deconfinement phase transition 
of either heavy static quarks with zero baryon chemical potential or quarks with finite baryon chemical potential, aligning with the results in our previous work~\cite{Li:2021qer}.
The baryon chemical potential exerts negligible influence on the power spectra of GWs generated by confinement-deconfinement phase transitions. This is attributed to the close proximity of the phase transition temperatures in both the finite baryon chemical potential and zero chemical potential scenarios.
Furthermore, it is worth noting that, based on the Bayesian method, our analysis has revealed that the phase transition of a pure gluon plasma cannot be regarded as the potential source of the NANOGrav signal, and the non-runaway bubble dynamics model is effectively ruled out by the NANOGrav data at a significance level of $5\sigma$.
Secondly, we obtain a remarkably distinct result compared to the previous work, i.e., the bubble dynamics in the phase transition should be characterized by the Jouguet detonation rather than a non-runaway scenario. 
Thirdly, our results provide some insights into the applicability of holographic descriptions of the quark phase transitions, 
in which the models of hard and soft walls relate to the introduction of sharp and smooth cutoffs respectively. According to our analysis, the holographic description using the hard wall model for the quark phase transition is more likely to be consistent with PTA data as compared to that using the soft wall model.

In this work, we only examine the first-order confinement-deconfinement phase transitions. However, it is also worth considering other types of the first-order phase transition at the QCD scale, such as the chiral symmetry-breaking~\cite{Reya:1974gk}.
Moreover, relying solely on PTA observations might not be sufficient to distinguish the various details of different first-order phase transitions. Combining additional constraints~\cite{Bai:2021ibt} related to first-order phase transitions will help us gain further insights into the processes of cosmic phase transitions. We leave these aspects for future research.

\emph{Acknowledgments}
ZCC is supported by the National Natural Science Foundation of China (Grant No.~12247176 and No.~12247112) and the China Postdoctoral Science Foundation Fellowship No. 2022M710429. 
SLL is supported by the National Natural Science Foundation of China (Grant No.~12105098, No.~11947216, No.~12005059) and the Natural Science Foundation of Hunan Province (Grant No.~2022JJ40264).
HWY and PXW are supported by the National Key Research and Development Program of China Grant No.~2020YFC2201502, and by the National Natural Science Foundation of China under Grants No.~12275080 and No.~12075084.

\bibliographystyle{apsrev4-1}
\bibliography{ref}

\end{document}